\documentclass[12pt]{article}
\begin{document}
\begin{center}
\Large{\textbf{Dynamical probability, particle trajectories and completion of 
traditional quantum mechanics}}

\vspace{.2in}
\large{Tulsi Dass}

Chennai Mathematical Institute, 92 G. N. Chetty Road, \mbox{T. Nagar}, Chennai 
600017, India
\end{center}

\vspace{.2in}

\noindent
\textbf{Abstract.} Maintaining the position that the wave function $\psi$ 
provides a 
complete description of state, the traditional formalism of quantum 
mechanics is  augmented by introducing continuous trajectories for particles 
which are sample paths of a stochastic process  determined 
(including the underlying probability space) by $\psi$. In the resulting 
formalism, problems relating to measurements and objective reality are 
solved as in Bohmian mechanics (without sharing its weak points). The 
pitfalls of Nelson's stochastic mechanics are also avoided.

\newpage
\noindent
\emph{We have yet to understand} \\
\emph{How rich} \\
\emph{And deep} \\
\emph{The quantity $\psi$ is.}

\vspace{.15in}
\noindent
\section{Introduction} 
 This work concerns the following three  deficiencies of the 
traditional formalism of quantum mechanics (QM): failure in describing 
individual systems (it deals with ensembles), in giving a concrete description 
of systems as objectively real entities and in giving a satisfactory account of measurements free of the problem of macroscopic superpositions. 

As a remedy, a (minimal) completion of the formalism is proposed which is 
based on  the following simple idea: The wave 
function  $\psi(\mathbf{r},t)$ (of, say, a nonrelativistic spinless particle) 
defines not only a (time-dependent) probability measure on the configuration 
space $R^3$ [thus giving a parametrised family of probability spaces 
$\mathcal{M}_t = (R^3, \mathcal{B}(R^3), |\psi(\mathbf{r},t)|^2 d^3x)$ where 
$\mathcal{B}(R^3)$ is the Borel $\sigma$-algebra of $R^3$] but also a 
stochastic process \textbf{X}(t) [see Eq.(18) below] (say, for $t\geq0$) 
for 
which the underlying probability space is $\mathcal{M}_0$. The traditional 
formalism of QM is proposed to be supplemented with the following additional 
postulate :

\noindent
Postulate $\mathcal{P}$. The quantum mechanical  system 
whose dynamics is represented by the Schr$\ddot{o}$dinger 
equation for the wave function $\psi$ [Eq.(9) below] is a concrete 
physical entity --a particle-- having well defined position at 
every instant of time; the allowed configuration space trajectories 
for it are  the (continuous) sample paths of the stochastic process 
\textbf{X}(t) 
[labelled by the initial positions of  
the particle (points of $\mathcal{M}_0$)]. 

The theory permits the  experimental realization of these paths 
arbitrarily closely but not exactly 
(the situation in this respect is similar to that for the absolute zero 
of temperature). The formalism evolved will have the  good points of the 
de Broglie-Bohm theory or Bohmian mechanics (BM) (de Broglie 1926, 1927; 
Bohm 1952; Bohm and Vigier 1954; Bohm and Hiley 1993; Bell 1987; 
Holland 1993; Goldstein 1998) without sharing its 
weak points (and retaining 
all the good features of the traditional formalism) and also avoids the 
pitfalls of Nelson's stochastic mechanics (Nelson 1966,67; Guerra 1981; 
Nelson 1985; Blanchard et al 1987). (Both of these theories 
propose to do QM with particles moving along physically well defined 
trajectories.)

In section 2, we cover some background relating to the Hamilton-Jacobi theory 
and its probabilistic extension, de Broglie -Bohm theory and Nelson's 
stochastic mechanics. The main work mentioned above is presented in section 3. 
The last section contains some additional remarks.

\vspace{.15in}
\noindent
\section{Background}

\vspace{.12in}
\subsection{Hamilton-Jacobi theory and its probabilistic extension}

\vspace{.12in}
 We shall recall a few essential points relating to this theory (Guerra 1981; 
 Blanchard et al 1987; Holland 1993; Dass 2002). Given a classical system 
$\mathcal{S}$ with a Lagrangian $L(q,\dot{q},t)$ having configuration $q_0$ 
at the initial time $t_0$, the Hamilton-Jacobi function 
\begin{eqnarray}
S(q,t) = \int_{t_0,q_0}^{t,q} dt^{\prime} L(q(t^{\prime}),\dot{q}(t^{\prime}),
t^{\prime})
\end{eqnarray}
where the integration is along the physical trajectory between $q(t_0) =q_0$ 
and $q(t) = q$ (assumed, for simplicity, unique) satisfies the Hamilton-Jacobi 
equation 
\begin{eqnarray}
\frac{\partial S(q,t)}{\partial t} + H\left(q, \frac{\partial S(q,t)}
{\partial q},t\right) = 0
\end{eqnarray}
where $H(q,p,t)$ is the Hamiltonian. A solution $S(q,t)$ of Eq.(2) with the 
initial condition $S(q,t_0) = S_0(q)$, when supplemented with the initial 
condition 
\begin{eqnarray}
q(t_0) = q_0
\end{eqnarray}
can be used to obtain the (unique) dynamical trajectory (q(t),p(t)) in phase 
space given by the initial condition 
\begin{eqnarray*}
(q(t_0),p(t_0)) = (q_0, p_0) \hspace{.12in} \mathrm{where} \hspace{.12in} 
p_{0\alpha} = 
\left(\frac{\partial S_0(q)}{\partial q^{\alpha}} \right)_{q=q_0}
\end{eqnarray*}
as follows: Define momentum and velocity fields $p(q,t)$ and $v(q,t)$ by 
\begin{eqnarray}
p_{\alpha}(q,t) = \frac{\partial S(q,t)}{\partial q^{\alpha}}; \hspace{.15in} 
v^{\alpha}(q,t) = \left(\frac{\partial H(q,p,t)}{\partial p_{\alpha}}
\right)_{p=p(q,t)}.
\end{eqnarray}
The differential equation
\begin{eqnarray}
\dot{q}_{\alpha}(t) = v^{\alpha}(q(t),t)
\end{eqnarray}
with the initial condition (3) gives a unique solution $q(t)$. Finally 
$p_{\alpha}(t) = p_{\alpha}(q(t),t)$.

In this description of classical dynamics, the state at time t is given by 
$q^{\alpha}(t)$ and the field $S(q,t)$; their time evolution is governed by 
the equations (5) and (2). For a particle in potential $V(x)$, we have 
$H = (2m)^{-1}p^2 + V(x)$ and 
\begin{eqnarray}
v^j(x,t) = \frac{1}{m} \frac{\partial S(x,t)}{\partial x^j}; \hspace{.12in} 
j = 1,2,3.
\end{eqnarray}

If, instead of the condition (3), we are initially given a probability 
distribution $\rho (q,t_0) = \rho_0(q)$, we shall obtain, instead of the 
configuration space trajectories $q^{\alpha}(t)$, the probability density 
function $\rho (q,t)$ satisfying the continuity equation
\begin{eqnarray}
\frac{\partial \rho (q,t)}{\partial t} + \frac{\partial}{\partial q^{\alpha}}
[v^{\alpha}(q,t)\rho (q,t)] = 0. 
\end{eqnarray}
The state of the system at time t is now given by the pair of fields 
$\rho (q,t)$ and $S(q,t)$ whose evolution is governed by the equations (7) 
and (2).

A point worth emphasising is that, even in this probabilistic version of 
Hamilton-Jacobi theory, the concept of configuration space trajectories 
(which are particle trajectories if $\mathcal{S}$ is a system of particles) 
governed by Eq.(5) remains meaningful; we only do not have \emph{complete} 
information 
about them. Probabilistic questions relating to (families of) such 
trajectories can be answered in the formalism. Taking the system 
$\mathcal{S}$, for convenience, to be a nonrelativistic particle,
the trajectories are now the sample paths of the stochastic process 
$\mathbf{x}(t)$ governed by the stochastic differential equation (SDE) 
\begin{eqnarray}
dx^j(t) = \frac{1}{m}\left(\frac{\partial S(x,t)}{\partial x^j} 
\right)_{x = x(t)} dt.
\end{eqnarray} 
The underlying probability space for this process can be taken as 
$\mathcal{M}_0^{cl} = (R^3, \mathcal{B}(R^3), \rho_0(\mathbf{r})d^3x)$ 
(points of the sample space $R^3$ being possible initial positions of the 
particle). 

Note that, to have the probabilistic treatment of particle trajectories 
along the lines mentioned above, we do not need to compromise with the fact 
that the fields $\rho(\mathbf{r},t)$ and $S(\mathbf{r},t)$ constitute a 
complete description of state at time t. These functions, determined as 
solutions of Eqs.(7) and (2) with initial conditions $\rho(\mathbf{r},t_0) = 
\rho_0(\mathbf{r})$ and $ S(q,t_0) = S_0(q)$ determine the probabilty space 
$\mathcal{M}_0^{cl}$ 
as well as the SDE (8) (and, therefore, the family of sample paths of this SDE 
which are particle trajectories). In particular, it is \emph{not} true that, 
to have Eq.(8) in the formalism, the state at time t must be taken to be given 
by  $\rho(\mathbf{r},t), S(\mathbf{r},t)$ \emph{and} the trajectory function 
$\mathbf{r}(t)$.

In the limiting case $ \rho_0 (\mathbf{r}) = \delta (\mathbf{r} - \mathbf{r}_0) $, we 
recover the deterministic trajectories. If we insist on $\rho$ being a 
well behaved function, such trajectories can be realised arbitrarily closely 
but not exactly.

\vspace{.12in}
\noindent
\subsection{Bohmian mechanics}

\vspace{.12in}
Bohmian mechanics (BM) treats particles as concrete physical 
entities moving in a highly non-Newtonian manner along well defined 
trajectories. Each particle has associated with it a complex \emph{physical} 
field $\psi(\mathbf{r},t)$. The complete description of state of the system 
(particle plus its associated field) at time t is given by the field 
$\psi(\mathbf{r},t)$ and the trajectory function $\mathbf{r}(t)$. Temporal 
evolution of state is assumed to be given by the Schr$\ddot{o}$dinger 
equation
\begin{eqnarray}
i\hbar\frac{\partial \psi(\mathbf{r},t)}{\partial t} = [-\frac{\hbar^2}{2m}
\bigtriangledown^2 + V(\mathbf{r})] \psi(\mathbf{r},t)
\end{eqnarray}
and an equation identifying the velocity of the particle with the velocity 
of the Schr$\ddot{o}$dinger field $\psi(\mathbf{r},t)$ at its location:
\begin{eqnarray}
\frac{d \mathbf{r}(t)}{dt} = \mathbf{v}^{(\psi)}(\mathbf{r}(t),t) 
\end{eqnarray}
where
\begin{eqnarray} 
\mathbf{v}^{(\psi)}(\mathbf{r},t) =  \left( \frac{\mathbf{J}^{(\psi)}}
{\psi^*\psi} \right)(\mathbf{r},t) 
\equiv \frac{\hbar}{m} \left[\frac{Im(\psi^* \mathbf{\bigtriangledown} \psi)}
{\psi^* \psi } \right] (\mathbf{r},t).
\end{eqnarray}
Eq.(9) implies the well-known continuity equation
\begin{eqnarray}
\frac{\partial |\psi(\mathbf{r},t)|^2}{\partial t} + 
\mathbf{\bigtriangledown}.\mathbf{J}^{(\psi)}(\mathbf{r},t) = 0.
\end{eqnarray}
Note that Eqs. (9) and (10) constitute a deterministic system admitting, 
with initial conditions $\mathbf{r}(0) = \mathbf{r}_0, \psi(\mathbf{r},0) = 
\psi_0(\mathbf{r})$, a unique solution (valid in an appropriate domain).

In a statistical ensemble of particles having a common associated field $\psi$ 
(but possibly different initial positions), the probability density of 
position is assumed to be given by
\begin{eqnarray}
\rho(\mathbf{r},t) = |\psi(\mathbf{r},t)|^2.
\end{eqnarray}
In fact, it is adequate to assume Eq. (13) at a single instant, say, $t = 0$; 
Eq.(12) then guarantees its validity for all t. 

The last assumption [Eq. (13)] was criticised by Pauli and others on the 
ground that such a hypothesis is not appropriate for a theory aimed at 
giving a causal explanation of QM. To counter this criticism, Bohm and 
Vigier (1954) proposed a hydrodynamic model taking the wave function to 
represent  a conserved fluid with density $|\psi|^2$ and local stream velocity 
$\mathbf{v}^{(\psi)}$ [which, on writing $ \psi = \tilde{R}exp[i \tilde{S}/
\hbar]$, equals $(\mathbf{\bigtriangledown}\tilde{S})/m$]. The particle is 
treated as an inhomogeneity which moves with the local stream velocity. By 
introducing a hypothesis of a very irregular and effectively random 
fluctuation in the motion of the fluid, they were able to prove that an 
arbitrary probability density ultimately decays into $|\psi|^2$. (For a 
relatively recent treatment of this `quantum equilibrium', see D$\ddot{u}$rr 
et al 1992.)
Eq. (10) is now replaced by 
\begin{eqnarray}
\frac{d \mathbf{r}(t)}{dt} = \mathbf{v}^{(\psi)}(\mathbf{r}(t),t) + 
\mathbf{\xi}(t)
\end{eqnarray}
where $\mathbf{\xi}$ represents the chaotic contribution to the particle 
velocity. Apart from some qualitative statements about it, no detailed 
treatment of this quantity was given.

The theory does not have the traditional quantum mechanical observables in 
its initial formulation; they can, however, be introduced through analysis 
of experiments (D$\ddot{u}$rr et al 2003). For observers not having knowledge of precise 
initial positions of particles, BM can be shown to make the same predictions 
as the standard quantum theory.(For such observers, BM is effectively a 
hidden variable theory, the position variables of particles serving as 
the hidden variables.)

Major plus points of BM are: \\
(i) It gives the simplest and most appealing account of the wave 
function reduction. In the macroscopic superposition of possible outcomes 
resulting from von Neumann's treatment of measurement (von Neumann 1955, 
Dass 2005), the one term 
representing the `channel' in which the system trajectory happens to be 
is naturally picked up. \\
(ii) It permits additional insights into various quantum phenomena like 
barrier penetration, interference etc in terms of particle trajectories. \\
(iii) It makes quantum nonlocality manifest (by exhibiting the `quantum 
potential' term in the treatment of particle motion). \\
(iv) The theory describes quantum systems as concrete physical systems 
having objective reality.

The somewhat overenthusiastic support of Bell notwithstanding, the theory 
is not without some weak points; the major ones are: \\
(i) Serious problems in the relativistic domain and quantum field theory; \\
(ii) Rather `painful' treatment of spin in the nonrelativistic domain; \\
(iii) Some unappealing/unconvincing aspects of the physics of the 
$\psi$-field. (For example, it influences particle motion but is not 
influenced by it; there is no discussion of the energy, momentum etc of 
this field, etc.)

The troubles of BM basically arise from the fact that it tries to overdo a 
couple of (related) things: \\
(i) It assigns physical status to not only the trajectory functions 
$\mathbf{r}(t)$ but also to $\psi$ and has to struggle with problems 
resulting from it. \\
(ii) The complete specification of state at time t includes not only the 
wave function$\psi(\mathbf{r},t)$, but also the trajectory function 
$\mathbf{r}(t)$.This is analogous to including, in the probabilistic version 
of Hamilton-acobi theory, $\mathbf{r}(t)$ besides $\rho(\mathbf{r},t)$ and 
$S(\mathbf{r},t)$ in the definition of state at time t which, as we have 
seen above, is not necessary.\\
As a result, there is an internal tension and awkwardness/uncleanliness 
in the formalism. This, in the author's opinion, is at least partly 
responsible for the theory's problems in the relativistic domain. Only a 
theory having a reasonably clean formalism at a certain level can be 
expected to have a straightforward generalization to a higher level.

\vspace{.12in}
\noindent
\subsection{Nelson's stochastic mechanics}

\vspace{.12in}
 According to stochastic mechanics (SM), quantum effects in the dynamics 
of microscopic systems arise due to the influence of a background field. It 
is assumed that this field causes the particles to undergo stochastic 
motion (described in terms of a stochastic process $X_t$) which 
is (i) Markovian, (ii)diffusion, which is (iii) conservative 
(i.e., on the average, there is no energy transfer between the system and 
the background field). In this kinematic framework, the dynamics is given 
(for, say, a nonrelativistic particle) in terms of a stochastic version of 
Newton's second law which is either assumed or derived from a stochastic 
version of Hamilton's principle. From the equation for this law and the 
continuity equation for the probability density $\tilde{\rho}$ for the 
stochastic process $X_t$, it follows that the quantity
\begin{eqnarray}
\psi = \sqrt{\tilde{\rho}}exp[i \tilde{S}/\hbar]
\end{eqnarray}
[where $\tilde{S}$ is a quantity whose gradient (plus a known function of 
$\tilde{\rho}$) defines the drift term in the SDE for the stochastic process 
$X_t$] satisfies the traditional Schr$\ddot{o}$dinger 
equation.

Observations on the system are analysed in terms of quantities related to 
the diffusion process $X_t$; typically, these are random variables of the type 
$f(X_{t_1},...,X_{t_n})$. For position observables, the predictions of SM 
and the traditional QM are identical. Taking the view that, in principle, 
all measurements can be analysed in terms of position measurements, there 
is a broad agreement between the predictions of the two theories. 

Detailed investigations, however, show that SM predicts undesirable 
correlations between distant systems which made Nelson himself 
conclude (Nelson 1984) that `Markovian stochastic mechanics is untenable as 
a realistic physical theory'. 

Here is a simple argument why Markov processes cannot provide the 
appropriate framework for the description of dynamics of objects whose 
wave functions satisfy Schr$\ddot{o}$dinger equation. Such a formalism 
is `overconstrained' in the sense that it has not only the 
Schr$\ddot{o}$dinger evolution semigroup of quantum dynamics, but also 
the evolution semigroup of the Markov process $X_t$ (related to the 
Chapman-Kolmogorov equation). Normally the former does not imply the latter. 
[One simple way to see this is to note that Schr$\ddot{o}$dinger evolution 
 in general mixes the 
diagonal and off-diagonal terms in the density matrix $w(x,y) = 
\psi(x)\psi^*(y)$; the 
Markovian semigroup mixes only the diagonal terms.]

The idea of starting with a background field hypothesis and producing a 
quantity $\psi$ satisfying Schr$\ddot{o}$dinger equation is very appealing; 
however, to do justice to this theme and produce a viable formalism, one 
needs to operate in a framework broader than the Markovian one.

\vspace{.15in}
\noindent
\section{Particle trajectories in quantum mechanics}

\vspace{.12in}
We shall present the proposed completion of the traditional formalism of QM 
in the context of a nonrelativistic spinless particle whose wave function 
satisfies the Schr$\ddot{o}$dinger equation (9). Some remarks relating to the 
generalization to to other systems will be made at an appropriate stage. 

The last part of section 2.1 makes it almost inevitable to invoke the 
postulate $\mathcal{P}$ which we presently do and proceed to explore 
the consequences. 
Given a 
solution $\psi(\mathbf{r},t)$ of Eq.(9) [supplemented with the initial 
condition $\psi(\mathbf{r},0) = \psi_0(\mathbf{r})$], we introduce, in the 
probability space $\mathcal{M}_0$, a stochastic process 
$\mathbf{X}(t;[\psi_0])$ (the square bracket in $[\psi_0]$ reflects the 
fact that the objects $\mathbf{X}(.;.)$ are functionals of $\psi_0$) 
whose sample paths 
$ \mathbf{X}(t;[\psi_0], 
\mathbf{r}_0)$ (where $\mathbf{r}_0$ labels points of the configuration space 
$R^3$ considered at time t = 0) are supposed to represent particle 
trajectories. Two obvious conditions to impose on these functions are :
\begin{eqnarray}
 \mathbf{X}(0; [\psi_0], \mathbf{r}_0) = \mathbf{r}_0; 
\end{eqnarray}
and  $\mathbf{X}(t; [\psi_0], \mathbf{r}_0) = 0$ for those values of 
$\mathbf{r}_0$ for which $ \psi_0(\mathbf{r}_0) = 0$. \\
We shall often suppress the arguments $[\psi_0]$ and $\mathbf{r}_0$ and 
restore them whenever clarity demands it.

To ensure consistency with the traditional operator formalism, we demand 
that the expectation value of $\mathbf{X}(t)$ (in the context of the 
probability space $\mathcal{M}_0$) must be equal to the expectation value 
of the position  operator $\hat{\mathbf{X}}_H(t)$ [notation : the subscripts 
H  and S refer to Heisenberg and Schr$\dot{o}$dinger pictures (identified at 
time t= 0); 
$ <\mathbf{r}| \Psi_S(t)> = \psi_S(\mathbf{r},t)$; we shall often suppress 
the subscript S in $\psi_S(\mathbf{r},t)$] :
\begin{eqnarray}
E[X^j(t)] & = & \int_{R^3}X^j(t;[\psi_0],\mathbf{r}_0)
|\psi_0(\mathbf{r}_0)|^2 dx_0 \nonumber \\ 
& = & <\Psi_H|\hat{X}^j_H(t)|\Psi_H> \nonumber \\
& = & <\Psi_S(t)|\hat{X}^j_S|\Psi_S(t)> \nonumber \\
& = & \int_{R^3} x^j |\psi_S(x,t)|^2 dx \hspace{.15in} j = 1,2,3.
\end{eqnarray}

We next try to set up, for the process $\mathbf{X}(t)$, a stochastic 
differential equation (SDE) of the form
\begin{eqnarray}
dX^j(t) = V^j(\mathbf{X}(t),t)dt + d\zeta^j(t)
\end{eqnarray}
where the first term on the right (the traditional drift term) represents the 
mean motion during the time interval (t, t+dt) and the last term represents 
fluctuations. We impose the reasonable demand that the fluctuations 
average out to zero :
\begin{eqnarray}
E(d\zeta^j(t)) = 0.
\end{eqnarray}
This gives [recalling, for the intermediate steps, Eqs.(10) and (12)]
\begin{eqnarray*}
E(V^j(\mathbf{X}(t),t))dt &=& dt \int_{R^3} V^j(x,t)|\psi(x,t)|^2dx \\
                          &=& E(dX^j(t)) \\
&=& \int x^j [ |\psi(x, t+dt)|^2 - |\psi(x,t)|^2] dx \\
&=& dt \int x^j [-\mathbf{\bigtriangledown.J^{(\psi)}}(x,t)dx \\
&=& dt \int J^j(x,t)dx
\end{eqnarray*}
which in turn gives (upto an additive term of zero mean)
\begin{eqnarray}
V^j(\mathbf{r},t) = J^j(\mathbf{r},t)|\psi(\mathbf{r},t)|^{-2} 
= {v^{(\psi)}}^j(\mathbf{r},t).
\end{eqnarray}
We absorb the possible additive term with zero mean in the last term in 
Eq.(18). To do it transparently, we \emph{define} $\zeta^j$ as 
\begin{eqnarray}
\zeta^j(t;[\psi_0],\mathbf{r}_0) \equiv X^j(t;[\psi_0],\mathbf{r}_0) 
- x_0^j - \int_0^t dt^{\prime} {v^{(\psi)}}^j(\mathbf{X}(t^{\prime};
[\psi_0],\mathbf{r}_0), t^{\prime}).
\end{eqnarray}
With $V^j$ given by Eq.(20), Eq.(18) is now equivalent to Eq.(14) 
[making allowance for the fact that $\mathbf{\xi}(t)$ may be a generalized 
stochastic process (white noise, for example)]. We have, therefore, got the 
main desirable equation of BM essentially for free! 

In fact, we are in an even better position. To see this, it is useful to 
recall  the velocity path integral for the transition amplitude 
\mbox{(Kleinet, 2004, p 172):} 
\begin{eqnarray}
<\mathbf{x_b},t_b|\mathbf{x_a},t_a> = \int \mathcal{D}^3v 
\delta (\mathbf{x_b} - \mathbf{x_a} - \int_{t_a}^{t_b}dt \mathbf{v}(t)). 
\nonumber \\
exp\left\{ \frac{i}{\hbar}\int_{t_a}^{t_b}dt \left[\frac{m}{2}|\mathbf{v}|^2 - 
V\left(\mathbf{x_b} - \int_t^{t_b}dt \mathbf{v}(t)\right) \right] \right\}
\end{eqnarray}
where the path integrator $\mathcal{D}^3v $ satisfies the normalization 
condition 
\begin{eqnarray}
\int\mathcal{D}^3v exp\left[ \frac{i}{\hbar}\int_{t_a}^{t_b} dt 
\left(\frac{m}{2} |\mathbf{v}|^2 \right) \right] = 1.
\end{eqnarray}
If, in Eq.(23), we put $(m\hbar^{-1})^{1/2}\mathbf{v} = 
\tilde{\mathbf{v}}$, and define $\mathcal{D}^3\tilde{v}= \mathcal{D}^3v$,
the resulting equation will have no dimensional parameters. Accordingly, 
writing $\mathbf{\xi} = (\hbar m^{-1})^{1/2}\tilde{\mathbf{\xi}}$ and noting 
that the velocity path integral in Eq.(21) is an integral over velocity 
fluctuations of the particle, one expects $\tilde{\mathbf{\xi}}$ to be a 
quantity with standard normalization. The 
resulting equation is important enough to deserve a separate display :
\begin{eqnarray}
\frac{d\mathbf{X}(t)}{dt} = \mathbf{v}^{(\psi)}(\mathbf{X}(t),t) + 
\sqrt{\frac{\hbar}{m}}\tilde{\mathbf{\xi}}.
\end{eqnarray}
With $\hbar \neq 0$ (as is the case with the 
real world), the last term in Eq.(24) is insignificant for particles of 
large mass. This is quite consistent with observations.

Recalling Eq.(15), if the limits
\begin{eqnarray}
\rho(\mathbf{r},t) = lim_{\hbar \rightarrow 0} \tilde{\rho}(\mathbf{r},t); 
\hspace{.15in} S(\mathbf{r},t) = lim_{\hbar \rightarrow 0} \tilde{S}
(\mathbf{r},t)
\end{eqnarray}
exist (and are smooth functions), the real and imaginary parts of Eq.(9) 
go over, in the limit $\hbar \rightarrow 0$, to Eqs. (7) and (2). In this 
limit, Eq.(24) goes over to the special case of Eq.(5) corresponding to 
the velocity field of Eq.(6) [see Eq.(8)]. The limiting trajectories are smooth. 

The augmented formalism [to be referred to as CQM (Completed QM)] now has 
the plus points of BM given above. In 
particular, the explanation of wave function collapse can be given as, 
for example, in Ch.(6) of Bohm and Hiley [described briefly in the plus 
point (i) in section 2.1].

 To clear some lingering doubts as to why, in a formalism admitting 
trajectories (in configuration space), the traditional quantum mechanical 
observables (self adjoint operators) -- and not what Bell calls be-ables --
should be the proper choice for mathematical objects representing 
measurable quantities, note the following points: \\
(i) The trajectories in CQM are continuous but not smooth. There is, 
therefore, no concept of velocity (as a well defined function of time) at a  
general point of a trajectory. 
It follows that be-ables are not a natural choice of observables in 
this theory. \\
(ii) In CQM, trajectories are derived objects; the central objects are the 
wave functions which give a complete description of state. A straightforward 
logical analysis of measurements (Araki 1999, Dass 2002) on systems on 
which repeatable 
experiments can be performed (to facilitate frequency interpretation of 
probability), leads to the dual relationship between states and observables. 
When the description of state is given in terms of vectors in a Hilbert 
space (or, more generally, in terms of density operators), the proper 
choice of obsrvables, according to this analysis, is an appropriate 
family of self-adjoint operators. This is by no means inconsistent with the 
picture of particles (or more general systems) moving along continuous 
trajectories in configuration space.

\vspace{.15in}
\noindent
\section{Additional remarks}

\vspace{.12in}
\noindent
(i) Two points of contrast between the present work and Nelson's stochastic 
mechanics  are : \\
(a) Whereas, in SM, the stochastic process $X_t$ is defined on an abstract 
(or background) probability space, we employ only the dynamical probability 
spaces $\mathcal{M}_t$ provided by the wave function. (Hence the term 
`dynamical probability' in the title of the paper.) \\
(b) We do not make the Markovian assumption about th stochastic process 
$\mathbf{X}(t)$. 

\noindent
(ii) In SM, one has to distinguish between the forward differential 
$d^{(+)}X(t) = X(t+dt) - X(t)$ and the backward differential 
$ d^{(-)}X(t) = X(t) - X(t-dt)$ and correspondingly, in the drift term in 
the relevant SDEs, forward and backward velocities $v^{(\pm)}(x,t)$; the 
velocity  $v^{(\psi)}$ appearing in Eq.(23) is the mean of the two. 
Formally, therefore, our differential $dX^j(t)$ must be taken to correspond 
to the symmetric differential
\begin{eqnarray}
\circ X^j(t) = \frac{1}{2} [X^j(t+dt) - X^j(t-dt)].
\end{eqnarray}
In fact, trajectories in CQM are somewhat smoother objects than those in SM. 
For example, the calculational steps before Eq.(20) give the same result 
whether we use the forward, backward or symmetric differential.

\noindent
(iii) An equation analogous to (23) appears in Roncadelli [1993; Eq.(21)]. 
Instead 
of $v^{(\psi)}$, however, Roncadelli has the classical velocity term (6) 
(where S, in his work,  is a general solution of the Hamilton-Jacobi equation). 

\noindent
(iv) The functions $\zeta^j(t)$ (or, equivalently, $X^j(t)$) need to be 
determined completely (if necessary, by strengthening the postulate 
$\mathcal{P}$). The author hopes to come back to this point in a future 
publication. Meanwhile, we give here a relation worth putting on record.

From Eq.(17), one can write [for those $\mathbf{r}$ for which 
$\psi_0(\mathbf{r}) \neq 0$]
\begin{eqnarray}
X^j(t;[\psi_0],\mathbf{r}) = \frac{x^j |\psi(\mathbf{r},t)|^2}
{|\psi_0(\mathbf{r})|^2} + G^j(t;[\psi_0], \mathbf{r})
\end{eqnarray}
where [remembering Eq.(16)]
\begin{eqnarray}
G^j(0;..) = 0 \\
\int_{R^3} G^j(t;[\psi_0],x)|\psi_0(x)|^2dx = 0.
\end{eqnarray}
Comparing Eqs.(27) and (21), we see that the functions $G^j$ and $\zeta^j$ are
related. It is, therefore, adequate to determine one of them. 

\noindent
(v) The formalism of CQM, presented in section 3 in the context of a 
spinless nonrelativistic particle, admits straightforward generalization 
(with due alteration of details) to (at least) all systems with finite 
dimensional configuration spaces. In the statement of the postulate 
$\mathcal{P}$, the term `particle' is to be replaced by the relevant 
system and, in Eq.(24), the velocity $v^{(\psi)}$ is to be replaced by 
the velocity of the $\psi$-field appearing in the relevant continuity 
equation replacing Eq.(12). [See Eq.(7) for the form of the general 
continuity equation.] Whereas the wave function $\psi$ describes an ensemble, 
the trajectory functions [ the quantities $X^j(t;[\psi_0], \mathbf{r}_0)$ 
and their analogues in the more general case] describe, for a given $\psi$, 
the configuration space trajectories of individual systems in the ensemble.

\vspace{.15in}
\textsl{The author would like to thank the Chennai Mathematical Institute and 
C.S. Seshadri for hospitality and the Institute of Mathematical Sciences, 
Chennai and R. Balasubramanian for providing library and other facilities.}

\vspace{.15in}
\begin{description}
\item Araki H 1999 \textsl{Mathematical Theory of Quantum Fields} (Oxford 
University Press)
\item Bell J S 1987 \textsl{Speakable and Unspeakable in Quantum Mechanics} 
(Cambridge University Press)
\item Blanchard Ph et al 1987 \textsl{`Mathematical and physical aspects of 
stochastic mechanics', Lecture notes in physics, vol 281}(Berlin : Springer-
Verlag)
\item  Bohm D 1952 \textsl{Phys. Rev.} \textbf{85},166, 180  
\item Bohm D and  Hiley B J 1993 \textsl{The Undivided Universe: An 
Ontological Interpretation of Quantum theory} (ondon: Routledge,Chapman and 
Hall)
\item Bohm D and  Vigier J P 1954 Phys. Rev. \textbf{96}, 208 
\item de Broglie L 1926 \textsl{Compt. rend.} \textbf{183}, 447 
\item de Broglie L 1927 \textsl{Compt.rend.}\textbf{184}, 273; \textbf{185}, 
380 
\item Dass T 2002 \textsl{Towards an Autonomous Formalism for Quantum 
Mechanics}, arxiv: quant-ph/0207104.
\item Dass T 2005 \textsl{Measurements and decoherence}, 
arxiv: quant-ph/0505070
\item D$\ddot{u}$rr D et al 1992 \textsl{J. Stat. Phys.} \textbf{67}, 843;
 arxiv: quant-ph/0308039
\item D$\ddot{u}$rr D et al 2003, arxiv: quant-ph/0308038 
\item Goldstein S 1998 \textsl{Physics Today} \textbf{51}, No. 3, 42; 
No. 4, 38
\item Guerra F 1981 \textsl{Phys. Reports} \textbf{77}, 263 
\item Holland P R 1993 \textsl{The Quantum Theory of Motion}(Cambridge 
University Press)
\item Nelson E 1966 \textsl{Phys. Rev.} \textbf{150}, 1079 
\item Nelson E 1967 \textsl{Dynamical Theories ofBrownian Motion} (Princeton 
University Press)
\item Nelson E 1984 \textsl{Physica A} \textbf{124}, 509  
\item Nelson E 1985 \textsl{Quantum Fluctuations} (Princeton University 
Press)
\item Roncadelli M 1993 \textsl{J. Phys. A: Math. gen.} \textbf{26}, L949
\item J. von Neumann, \emph{Mathematical Foundations of QuantumMehanics} 
(Princeton University Press, 1955).

\end{description}
\end{document}